\pdfoutput=1
\documentclass[12pt]{article}
\usepackage{amsmath}
\usepackage{color}

\usepackage{amssymb}
\usepackage{graphicx}
\usepackage{adjustbox}
\usepackage{multirow}
\usepackage{booktabs,caption}
\usepackage[flushleft]{threeparttable}
\usepackage{float}
\newcommand{\be}{\begin{equation}}
\newcommand{\ee}{\end{equation}}
\newcommand{\bea}{\begin{eqnarray}}
\newcommand{\eea}{\end{eqnarray}}

\catcode`@=12

\begin{document}
\thispagestyle{empty}
\begin{center}
{\Large\bf
{Gravitational Wave Signatures from Domain Wall and Strong First-Order Phase Transitions in a Two Complex Scalar extension of the Standard Model}}\\
\vspace{1cm}
{{\bf Avik Paul}\footnote{email: avik.paul@saha.ac.in},
{{\bf Upala Mukhopadhyay}\footnote{email: upala.mukhopadhyay@saha.ac.in }},
{\bf Debasish Majumdar}\footnote{email: debasish.majumdar@saha.ac.in}}\\
\vspace{0.25cm}
{\normalsize \it Astroparticle Physics and Cosmology Division,}\\
{\normalsize \it Saha Institute of Nuclear Physics, HBNI,} \\
{\normalsize \it 1/AF Bidhannagar, Kolkata 700064, India}\\
\vspace{1cm}
\end{center}
\begin{abstract}
We consider a simple extension of Standard Model by adding two complex singlet scalars with a $\rm{U}\left(1\right)$ symmetry. A discrete $\mathcal{Z}_2 \times \mathcal{Z}^{\prime}_2$ symmetry is imposed in the model and the added scalars acquire a non zero vacuum expectation value (VEV) when the imposed symmetry is broken spontaneously. The real (CP even) parts of the complex scalars mix with the SM Higgs and give three physical mass eigenstates. One of these physical mass eigenstates is attributed to the SM like Higgs boson with mass 125.09 GeV. In the present scenario, domain walls are formed in the early Universe due to the breaking of discrete $\mathcal{Z}_2 \times \mathcal{Z}^{\prime}_2$ symmetry. In order to ensure the unstability of the domain wall this discrete symmetry is also explicitly broken by adding a bias potential to the Lagrangian. The unstable annihilating domain walls produce a significant amount of gravitational waves (GWs). In addition, we also explore the possibility of the production of GW emission from the strong first-order phase transition. We calculate the intensities and frequencies of each of such gravitational waves originating from two different phenomena of the early Universe namely annihilating domain walls and strong first-order phase transition. Finally, we investigate the 
observational signatures from these GWs at the future GW detectors such as ALIA, BBO, DECIGO, LISA, TianQin, Taiji, aLIGO, aLIGO+ and pulsar timing arrays such as SKA, IPTA, EPTA, PPTA, NANOGrav11 and NANOGrav12.5. 
\end{abstract}
\newpage
\section{Introduction}

The progress of the detection of gravitational wave (GW) event from a binary black hole merger confirmed by the Laser Interferometer Gravitational-Wave Observatory (LIGO) group \cite{LIGOScientific:2016,{LIGOScientific:2017}} results in a remarkable enhancement in the physics of cosmology and astrophysics. The GWs from the early Universe can provide significant information related to the high energy physics phenomena in early Universe as the GWs after the production propagate without suffering any interaction and conserve almost all necessary physics information \cite{Maggiore:1999vm}. The production mechanisms of primordial GWs are associated with various cosmological sources such as inflationary quantum fluctuations \cite{Starobinsky:1979ty}, preheating after inflation \cite{Khlebnikov:1997di}, strong first-order phase transitions in the early Universe \cite{Witten:1984rs, {Hogan:1986qda}}, topological defects of the domain walls, cosmic strings \cite{Caldwell:1991jj}-\cite{Vilenkin:2000} etc. In this work, we explore two different production mechanisms of GW namely annihilation of cosmic domain walls (DWs) and strong first-order phase transition (SFOPT), in a simple two complex scalar extended Standard Model (SM). 

Domain walls, the two-dimensional surface-like topological defects are originated when a discrete symmetry is spontaneously broken \cite{Polturak:2019hpo}. But generally the formation of domain walls appears to contradict some basic wisdom in cosmology \cite{Zeldovich:1974uw} because energy density of the domain walls affects the total energy density of the Universe and leads to a rapid expansion of the Universe which is disfavoured by the present observational results \cite{Saikawa:2017hiv}. One can solve this domain wall problem by considering the unstable domain walls which collapse before they would overclose the Universe \cite{Vilenkin:1981zs, {Larsson:1996sp}}. In a theory, the unstability of domain walls can be established by considering that the discrete symmetry is approximate and it is broken explicitly \cite{Saikawa:2017hiv}. After the formation of unstable domain walls they collide or annihilate and emit gravitational waves. Such GWs can contribute as a stochastic background of GW in the present Universe. Different types of particle physics models are considered in Refs. \cite{Saikawa:2017hiv, {Zhou:2020ojf}, {Kadota:2015dza}} for the production of GWs from domain walls.

On the other hand, GWs can also be emitted from strong first-order phase transitions. Initially, at high temperatures, the Universe is in a false vacuum state but gradually as the temperature of the Universe decreases as it expands and evolves, it shifts to the true vacuum state through the process of tunnelling. The process of tunnelling leads to the first-order phase transition via bubble nucleation. These bubbles could be of different sizes. The bubbles that are large enough for avoiding collapse will expand, collide and coalesce which distorting the spherical symmetry of the bubbles and eventually leads to phase transition and emission of GWs as a result. 
In this scenario, the GWs are produced via three mechanisms such as (i) bubble collisions \cite{Kosowsky:1992a}-\cite{Caprini:2008}, (ii) sound waves induced by the bubbles running through the cosmic plasma \cite{Hindmarsh:2014}-\cite{Hindmarsh:2015} and (iii) turbulence induced by the bubble expansions in the cosmic plasma \cite{Caprini:2006}-\cite{Caprini:2009}.

Although the first-order electroweak (EW) phase transition (PT) would have been possible in the framework of SM of particle physics through the Higgs mechanism but with the observed Higgs mass of 125.09 GeV \cite{Kajantie:1996}-\cite{Csikor:1999} the transition is a smooth cross-over and not a first-order one. However, in literature there are ample references where the authors have shown that SFOPT can be realised by simple extension of SM \cite{Mazumdar:2018dfl} -\cite{Huang:2017rzf}. In the present work however, we primarily explore GW emission from the annihilation of domain walls by proposing a simple extension of SM with two complex scalars. In addition we also furnish the GW emission from SFOPT within the same framework of our proposed model. Domain wall is formed when a discrete symmetry is spontaneously broken. It appears at the boundary of two domains that is produced following such spontaneous breaking of discrete symmetry. In this regard, as mentioned, we extend the SM by adding two extra complex scalars and impose a discrete $\mathcal{Z}_2 \times \mathcal{Z}^{\prime}_2$ symmetry to the model. The two complex scalars acquire non zero vacuum expectation values (VEVs) when the imposed discrete symmetry is spontaneously broken. The real parts of the complex scalars mix with the SM Higgs and give three physical mass eigenstates. One of the components of the physical scalars attributes to the SM like Higgs boson with mass 125.09 GeV. As the discrete symmetry in this framework is broken spontaneously, domain walls are formed which is made unstable by adopting a bias term that lifts the degenerate vacua. In Ref. \cite{{Zhou:2020ojf}} the authors have considered single scalar extension of SM with a $\mathcal{Z}_3$ symmetry and subsequent production of GWs from domain wall. In this work however we approach the formation and collapse of domain walls by proposing two scalars (complex) extension of SM with a $\mathcal{Z}_2 \times \mathcal{Z}_2^\prime$ symmetry. In addition we also demonstrate that by the extension of the two new complex scalars the effective degrees of freedom of the model increase by four units hence SFOPT can be possible in the model.

In Refs. \cite{Kozaczuk:2014kva}-\cite{Huang:2017rzf} the authors have proposed different particle physics models for the production of GWs from first-order phase transition but variance from their approaches in this work, we consider an extension of SM with two complex scalars in such a way that it can provide GW production from SFOPT as well as the GW production from unstable domain walls. 
We constrain the model parameters by using some theoretical and experimental constraints such as vacuum stability, perturbativity and the collider bounds. We choose five benchmark points (BPs) from the allowed model parameter space to calculate the GW intensity and peak frequency of GW for both the cases namely domain wall and SFOPT. We also investigate the detectability of such GWs at the future space-based detectors such as ALIA \cite{Gong:2015}, BBO \cite{Harry:2006}, DECIGO \cite{Seto:2001}, LISA \cite{Caprini:2016}, TianQin \cite{Luo:2015ght}, Taiji \cite{Guo:2018npi}, ground-based detectors such as aLIGO \cite{Harry:2010}, aLIGO+ \cite{Harry:2010} as well as low-frequency pulsar timing arrays (PTAs)\cite{pta} such as SKA \cite{ska1}-\cite{ska3}, IPTA \cite{ipta1}-\cite{ipta4}, EPTA \cite{epta1}-\cite{epta3}, PPTA \cite{ppta1, {ppta2}}, NANOGrav11 \cite{nano1}-\cite{nano4} and NANOGrav12.5 \cite{Arzoumanian:2020vkk,{DeLuca:2020agl}}.     

The paper is organised as follows. In section 2, we briefly describe our two complex scalars extended SM and also derive the necessary relations between model parameters. In section 3, we discuss some bounds which we have used to constrain the model parameter space. The formation of unstable domain walls and the calculations of intensities of the resulting GW emission for this extended model are presented in section 4. In section 5, we furnish the finite temperature effective potential to study the first-order phase transition properties in the framework of our model. The possible production mechanisms of GWs from the strong first-order phase transitions are also explored in this section. Then in Section 6, we calculate the intensity and frequency of gravitational waves produced from both the domain walls and strong first-order phase transitions. In this section, we also discuss the detectability of such GWs at the pulsar timing arrays and future GW detectors. Finally, we summarise our work and give some concluding remarks in section 7.

\section{The Model}
In this work, we extend the Standard Model of particle physics by adding two complex scalar singlets $S_1$ and $S_2$. A discrete $\mathcal{Z}_2 \times \mathcal{Z}^{\prime}_2$ symmetry is imposed in the model where under $\mathcal{Z}_2$ symmetry $S_1$ transforms as $-S_1$ and under $\mathcal{Z}^{\prime}_2$ symmetry $S_2$ transforms as $-S_2$. Thus under $\mathcal{Z}_2 \times \mathcal{Z}^{\prime}_2$ symmetry $S_1$ and $S_2$ are $\left(-1,1\right)$  and $\left(1,-1\right)$ respectively. The complex scalars $S_1$ and $S_2$ acquire non zero vacuum expectation values when the imposed $\mathcal{Z}_2$ and $\mathcal{Z}^{\prime}_2$ symmetries are broken spontaneously. The real parts (CP even) of the complex scalars mix with the SM Higgs and give three physical mass eigenstates. Here, we consider one of the physical mass eigenstates behaves as the SM Higgs boson with mass 125.09 GeV \cite{Patrignani:2016xqp}.  
 
The most general renormalisable scalar potential under the $\mathcal{Z}_2 \times \mathcal{Z}^{\prime}_2$ symmetry with $\rm{U}\left(1\right)$ symmetry can be written as
\begin{equation}\label{eq:1}
\begin{aligned}
V=\mu_H^2 \left(H^\dagger H\right) + \mu_{S_1}^2 \left(S_1^{*}S_1\right) + \mu_{S_2}^2 \left(S_2^{*}S_2\right) + \lambda_H \left(H^\dagger H \right)^2 + \lambda_{S_1} \left(S_1^{*}S_1 \right)^2 + \lambda_{S_2} \left(S_2^{*}S_2 \right)^2 \\+ \lambda_{HS_1} \left(H^\dagger H \right) \left(S_1^{*}S_1 \right) + \lambda_{HS_2} \left(H^\dagger H \right) \left(S_2^{*}S_2 \right) + \lambda_{S_1 S_2} \left(S_1^{*}S_1 \right) \left(S_2^{*}S_2\right) \hspace{2.7cm}\\+ g_{S_1 S_2} \left(S_1^{*}S_2 \right) \left(S_2^{*}S_1\right) + \dfrac{\kappa_{S_1S_2}}{2} \left[\left(S_1^{*}S_2 \right)^2+\left(S_2^{*}S_1 \right)^2 \right]\,\,,
\hspace{5.4cm}
\end{aligned}
\end{equation}
where $H$, $S_1$ and $S_2$ are the SM Higgs doublet and the two complex scalar singlets respectively. In the potential (Eq. \ref{eq:1}) we consider the mass parameters and all the couplings are real.

After spontaneous breaking of $\rm{SU}\left(2\right)_L \times \rm{U}\left(1\right)_Y$ symmetry SM Higgs field $H$ acquires a non zero VEV $v=$246.22 GeV and due to the spontaneous breaking of $\mathcal{Z}_2$ and $\mathcal{Z}^{\prime}_2$  symmetries the complex scalars $S_1$ and $S_2$ gain non zero VEVs $v_1$ and $v_2$ respectively ($v_1$ and $v_2$ are VEVs for $S_1$ and $S_2$ respectively). Therefore SM Higgs and the additional complex scalar fields can be represented as
\begin{equation}\label{eq:2}
H=\left(
\begin{array}{c}
 G^+\\
 \dfrac{1}{\sqrt{2}} \left(v+h+iG^0\right)\\
\end{array}
\right), \hspace{1mm} S_1=\dfrac{1}{\sqrt{2}}\left(v_1 + s_1 + i \chi_1 \right)
, \hspace{1mm} S_2=\dfrac{1}{\sqrt{2}}\left(v_2 + s_2 + i \chi_2 \right)\,\,,
\end{equation}
where $h$ , $s_1$, $s_2$, $\chi_1$ and $\chi_2$ are the unphysical scalars.
In Eq. \ref{eq:2}, $G^+$, $G_0$ are the charged and the neutral Goldstone bosons after the spontaneous symmetry breaking (SSB). 
We minimise the scalar potential (Eq. \ref{eq:1}) using the following conditions
\begin{equation}\label{eq:3}
\dfrac{\partial V}{\partial h}\Big|_{h=G^0=G^+=s_1=s_2=\chi_1=\chi_2=0}=\dfrac{\partial V}{\partial s_1}\Big|_{h=G^0=G^+=s_1=s_2=\chi_1=\chi_2=0}=\dfrac{\partial V}{\partial s_2}\Big|_{h=G^0=G^+=s_1=s_2=\chi_1=\chi_2=0}=0\,\,.
\end{equation}
The minimisation conditions lead us to the following tadpole relations
\begin{equation}\label{eq:4}
\begin{aligned}
\mu_H^2+\dfrac{1}{2}\lambda_{HS_1}v_1^2+\dfrac{1}{2}\lambda_{HS_2}v_2^2+\lambda_H v^2=0\,\,,\\
\mu_{S_1}^2+\dfrac{1}{2}\lambda_{HS_1}v^2+\dfrac{1}{2}\lambda_{S_1S_2}v_2^2+\dfrac{1}{2}g_{S_1S_2}v_2^2+\dfrac{1}{2}\kappa_{S_1S_2}v_2^2+\lambda_{S_1} v_1^2=0\,\,,\\
\mu_{S_2}^2+\dfrac{1}{2}\lambda_{HS_2}v^2+\dfrac{1}{2}\lambda_{S_1S_2}v_1^2+\dfrac{1}{2}g_{S_1S_2}v_1^2+\dfrac{1}{2}\kappa_{S_1S_2}v_1^2+\lambda_{S_2} v_2^2=0\,\,.
\end{aligned}
\end{equation}
By evaluating the second-order derivatives with respect to the associated fields masses of the scalars are obtained as 
\begin{equation}\label{eq:5}
m_{h}^2=\dfrac{\partial ^2V}{\partial {h}^2}=\mu_H^2+\dfrac{1}{2}\lambda_{HS_1}v_1^2+\dfrac{1}{2}\lambda_{HS_2}v_2^2+3\lambda_H v^2\,\,,
\end{equation}

\begin{equation}\label{eq:6}
m_{s_1}^2=\dfrac{\partial ^2V}{\partial {s_1}^2}=\mu_{S_1}^2+\dfrac{1}{2}\lambda_{HS_1}v^2+\dfrac{1}{2}\lambda_{S_1S_2}v_2^2+\dfrac{1}{2}g_{S_1S_2}v_2^2+\dfrac{1}{2}\kappa_{S_1S_2}v_2^2+3\lambda_{S_1} v_1^2\,\,,
\end{equation}

\begin{equation}\label{eq:7}
m_{s_2}^2=\dfrac{\partial ^2V}{\partial {s_2}^2}=\mu_{S_2}^2+\dfrac{1}{2}\lambda_{HS_2}v^2+\dfrac{1}{2}\lambda_{S_1S_2}v_1^2+\dfrac{1}{2}g_{S_1S_2}v_1^2+\dfrac{1}{2}\kappa_{S_1S_2}v_1^2+3\lambda_{S_2} v_2^2\,\,,
\end{equation}

\begin{equation}\label{eq:8}
m_{hs_1}^2=\dfrac{\partial ^2V}{\partial {h}\partial {s_1}}=\lambda_{HS_1}v v_1\,\,,
\end{equation}

\begin{equation}\label{eq:9}
m_{hs_2}^2=\dfrac{\partial ^2V}{\partial {h}\partial {s_2}}=\lambda_{HS_2}v v_2\,\,,
\end{equation}

\begin{equation}\label{eq:10}
m_{s_1 s_2}^2=\dfrac{\partial ^2V}{\partial {s_1}\partial {s_2}}=\left(\lambda_{S_1 S_2}+g_{S_1S_2}+\kappa_{S_1S_2}\right)v_1 v_2\,\,,
\end{equation}

\begin{equation}\label{eq:11}
m_{G^\pm}^2=\dfrac{\partial ^2V}{\partial {G^{\pm}}^2}=2\mu_H^2+2\lambda_H v^2+\lambda_{HS_1}v_1^2+\lambda_{HS_2}v_2^2\,\,,
\end{equation}

\begin{equation}\label{eq:12}
m_{G^0}^2=\dfrac{\partial ^2V}{\partial {G^0}^2}=\mu_H^2+\lambda_H v^2+\dfrac{1}{2}\lambda_{HS_1}v_1^2+\dfrac{1}{2}\lambda_{HS_2}v_2^2\,\,,
\end{equation}

\begin{equation}\label{eq:13}
m_{\chi_1}^2=\dfrac{\partial ^2V}{\partial {\chi_1}^2}=\mu_{S_1}^2+\dfrac{1}{2}\lambda_{HS_1}v^2+\dfrac{1}{2}\lambda_{S_1S_2}v_2^2+\dfrac{1}{2}g_{S_1S_2}v_2^2-\dfrac{1}{2}\kappa_{S_1S_2}v_2^2+\lambda_{S_1} v_1^2\,\,,
\end{equation}

\begin{equation}\label{eq:14}
m_{\chi_2}^2=\dfrac{\partial ^2V}{\partial {\chi_2}^2}=\mu_{S_2}^2+\dfrac{1}{2}\lambda_{HS_2}v^2+\dfrac{1}{2}\lambda_{S_1S_2}v_1^2+\dfrac{1}{2}g_{S_1S_2}v_1^2-\dfrac{1}{2}\kappa_{S_1S_2}v_1^2+\lambda_{S_2} v_2^2\,\,,
\end{equation}

\begin{equation}\label{eq:15}
m_{\chi_1 \chi_2}^2=\dfrac{\partial ^2V}{\partial {\chi_1}\partial{\chi_2}}=\kappa_{S_1 S_2} v_1 v_2\,\,.
\end{equation}
The real parts of the complex scalars in Eq. \ref{eq:2} mix with the neutral component of the SM Higgs doublet $h$. Considering the $3\times 3$ mass matrix in the $h-s_1-s_2$ basis we diagonalise the mass matrix by a unitary transformation $U\left(\theta_{12},\theta_{13},\theta_{23}\right)$ to obtain three physical mass eigenstates $h_1,\hspace{1mm}h_2$ and $\hspace{1mm}h_3$ in terms of the old basis $h-s_1-s_2$ and the three mixing angles $\theta_{12},\hspace{1mm}\theta_{13},\hspace{1mm}\theta_{23}$. The unitary transformation $U\left(\theta_{12},\theta_{13},\theta_{23}\right)$ has the following form 
\begin{equation}\label{eq:16}
\left(
\begin{array}{c}
 h_1 \\
 h_2  \\
 h_3  \\
\end{array}
\right)=U\left(\theta_{12},\theta_{13},\theta_{23}\right)\left(
\begin{array}{c}
 h \\
 s_1 \\
 s_2 \\
\end{array}
\right)\,\,,
\end{equation}
where $U\left(\theta_{12},\theta_{13},\theta_{23}\right)$ is the standard Pontecorvo-Maki-Nakagawa-Sakata (PMNS) \cite{DuttaBanik:2016jzv} matrix with the three mixing angles $\theta_{12},\hspace{1mm}\theta_{13},\hspace{1mm}\theta_{23}$ and  complex phase $\delta=0$. The matrix $U$ can be written as 
\begin{equation}\label{eq:18}
\footnotesize
\begin{aligned}
U=
\left(
\begin{array}{ccc}
U_{11} & U_{12} & U_{13}\\
 U_{21} & U_{22} & U_{23}\\
 U_{31} & U_{32} &U_{33}\\
\end{array}
\right)\hspace{12.22cm}\\=
\left(
\begin{array}{ccc}
 \cos\theta_{12} \cos\theta_{13} & \sin\theta_{12}\cos\theta_{13} & \sin\theta_{13}\\
 -\sin\theta_{12} \cos\theta_{23} -\sin\theta_{13} \sin\theta_{23}\cos\theta_{12} & \cos\theta_{23} \cos\theta_{12}- \sin\theta_{12}\sin\theta_{13}\sin\theta_{23}& \sin\theta_{23}\cos\theta_{13}\\
 \sin\theta_{12}\sin\theta_{23}-\sin\theta_{13}\cos\theta_{12}\cos\theta_{23} & -\sin\theta_{23}\cos\theta_{12}-\sin\theta_{12}\sin\theta_{13}\cos\theta_{23} &\cos\theta_{13}\cos\theta_{23}\\
\end{array}
\right)\,\,,
\end{aligned}
\end{equation}
so that Eq. \ref{eq:16} takes the form
\begin{equation}\label{eq:17}
\left(
\begin{array}{c}
 h_1 \\
 h_2 \\
 h_3 \\
\end{array}
\right)=
\left(
\begin{array}{ccc}
U_{11} & U_{12} & U_{13}\\
 U_{21} & U_{22} & U_{23}\\
 U_{31} & U_{32} &U_{33}\\
\end{array}
\right)\left(
\begin{array}{c}
 h \\
 s_1 \\
 s_2 \\
\end{array}
\right)\,\,.
\end{equation}
The three physical mass eigenstates $h_1$, $h_2$ and $h_3$ are therefore
\begin{equation}\label{eq:19}
h_1=U_{11}h+U_{12}s_1+U_{13}s_2, \hspace {1mm} h_2=U_{21}h+U_{22}s_1+U_{23}s_2, \hspace{1mm} h_3=U_{31}h+U_{32}s_1+U_{33}s_2\,\,.
\end{equation}
In this work we consider $h_1$ to be the SM like Higgs boson with mass $m_{h_1}=125.09$ GeV.
Using the minimisation conditions in Eq. \ref{eq:4}, one can obtain the following relations   
\begin{equation}\label{eq:20}
\lambda_H =\dfrac{m_{h}^2}{2v^2},\hspace{1mm} \lambda_{S_1} =\dfrac{m_{s_1}^2}{2v_1^2},\hspace{1mm}\lambda_{S_2} =\dfrac{m_{s_2}^2}{2v_2^2}\,\,.
\end{equation}

The model parameters ($\mu_H$, $\mu_{S_1}$, $\mu_{S_2}$, $\lambda_H$, $\lambda_{S_1}$, $\lambda_{S_2}$, $\lambda_{HS_1}$, $\lambda_{HS_2}$, $\lambda_{S_1 S_2}$) in Eq. \ref{eq:1} can be expressed in terms of the physical masses of the scalar particles $m_{h_1}$, $m_{h_2}$, $m_{h_3}$, VEVs of the scalar particles $v$, $v_1$, $v_2$ and by the three mixing angles $\theta_{12}$, $\theta_{13}$ and $\theta_{23}$. Thus we take $m_{h_1}$, $m_{h_2}$, $m_{h_3}$, $v$, $v_1$, $v_2$, $\theta_{12}$, $\theta_{13}$, $\theta_{23}$, $g_{S_1 S_2}$, and $\kappa_{S_1S_2}$ as the input parameters of the model and the benchmark points (BPs) are chosen for different sets of values of these parameters that satisfy all the constraints both theoretical and experimental described in the next section to explore the production of the gravitational waves from domain walls and strong first-order phase transition.

\section{Constraints}
In this section, we discuss some theoretical and experimental bounds which we have used to constrain the model parameter space.
\subsection{Theoretical Constraints}
\noindent \underline{\it Vacuum Stability}

The constraints on the parameter space from vacuum stability condition are obtained as \cite{Kannike:2012pe, {Paul:2019pgt}}
\begin{equation}\label{eq:22}
\lambda_H, \lambda_{S_1}, \lambda_{S_2}>0, \hspace{1mm} \lambda_{HS_1}+2\sqrt{\lambda_H \lambda_{S_1}}>0\,\,,
\end{equation}

\begin{equation}\label{eq:23}
\lambda_{HS_1}+g_{S_1 S_2}-\kappa_{S_1 S_2}+2\sqrt{\lambda_H \lambda_{S_1}}>0, \hspace{1mm}\lambda_{HS_2}+2\sqrt{\lambda_H \lambda_{S_2}}>0, \hspace{1mm} \lambda_{S_1S_2}+2\sqrt{\lambda_{S_1} \lambda_{S_2}}>0\,\,,
\end{equation}

\begin{equation}\label{eq:24}
\begin{aligned}
2\lambda_{HS_2}\sqrt{\lambda_{S_1}}+2\lambda_{S_1S_2}\sqrt{\lambda_H}+2\lambda_{HS_1}\sqrt{\lambda_{S_2}}\hspace{9cm}\\+2\left(2\sqrt{\lambda_H \lambda_{S_1} \lambda_{S_2}}+\sqrt{\left(\lambda_{HS_1}+2\sqrt{\lambda_H \lambda_{S_1}}\right) \left(\lambda_{HS_2}+2\sqrt{\lambda_H \lambda_{S_2}}\right)\left( \lambda_{S_1S_2}+2\sqrt{\lambda_{S_1} \lambda_{S_2}}\right)}\right)>0\,\,,
\end{aligned}
\end{equation}

\begin{equation}\label{eq:25}
\begin{aligned}
2\lambda_{HS_2}\sqrt{\lambda_{S_1}}+2\lambda_{S_1S_2}\sqrt{\lambda_H}+2\left(\lambda_{HS_1}+g_{S_1 S_2}-\kappa_{S_1 S_2}\right)\sqrt{\lambda_{S_2}}+2\Big(2\sqrt{\lambda_H \lambda_{S_1} \lambda_{S_2}} \hspace{3cm}\\+\sqrt{\left(\left(\lambda_{HS_1}+g_{S_1 S_2}-\kappa_{S_1 S_2}\right)+2\sqrt{\lambda_H \lambda_{S_1}}\right)\left(\lambda_{HS_2}+2\sqrt{\lambda_H \lambda_{S_2}}\right)\left( \lambda_{S_1S_2}+2\sqrt{\lambda_{S_1} \lambda_{S_2}}\right)}\Big)>0\,\,. \hspace{1cm}
\end{aligned}
\end{equation}

\noindent \underline{\it Perturbativity}

The perturbativity condition also sets another constraint on the quartic couplings in the tree-level potential (Eq. \ref{eq:1}) as \cite{Paul:2019pgt} $\left(\lambda_H,\hspace{1mm} \lambda_{S_1},\hspace{1mm} \lambda_{S_2},\hspace{1mm} \lambda_{HS_1},\hspace{1mm} \lambda_{HS_2},\hspace{1mm} \lambda_{S_1 S_2}, \hspace{1mm} g_{S_1 S_2} \hspace{1mm}{\rm{and}} \hspace{1mm}\kappa_{S_1S_2}\right) < 4\pi$.
\subsection{Experimental Constraints}
\noindent \underline{\it Collider Constraints}

In the present scenario, our model is extended by two new extra complex scalars and we expect that the newly added particles can affect the LHC collider physics phenomenology. In this work, we consider $h_1$ as the SM like Higgs boson with mass 125.09 GeV so we use the collider bounds to further constrain the model parameters.
The signal strength of the SM like Higgs $h_1$ is given by \cite{DuttaBanik:2016jzv}
\begin{equation}\label{eq:26}
R_1=U_{11}^4\dfrac{\Gamma ^{\text{SM}}}{\Gamma_1}\,\,,
\end{equation}
where $\Gamma ^{\text{SM}}$ is the total SM Higgs decay width. In  Eq. \ref{eq:26}, $\Gamma_1$ denotes the total decay width of SM like Higgs boson of mass 125.09 GeV which has the following form
\begin{equation}\label{eq:27}
\Gamma_1 =U_{11}^2\hspace{1mm} \Gamma ^{\text{SM}}+\Gamma_1^{\text{inv}}\,\,,
\end{equation}
where $\Gamma_1^{\text{inv}}$ refers to the invisible Higgs decay width. Here two invisible decay channels for $h_1$ can be possible. One of them is $\Gamma_1^{inv}\left(h_1\rightarrow h_2 h_2\right)$ (for  $m_{h_2}\leq m_{h_1}/2$) and the other is the $\Gamma_1^{inv}\left(h_1\rightarrow h_3 h_3\right)$ (for $m_{h_3}\leq m_{h_1}/2$). The total invisible Higgs decay width can be expressed by the sum of the decay widths of these two decay channels as
\begin{equation}\label{eq:28}
\Gamma_1^{inv}=\Gamma_1^{inv}\left(h_1\rightarrow h_2 h_2\right)+\Gamma_1^{inv}\left(h_1\rightarrow h_3 h_3\right)\,\,,
\end{equation}
The expressions for the decay channels are
 \begin{equation}\label{eq:29}
\Gamma_1^{inv}\left(h_1\rightarrow h_2 h_2\right)=\dfrac{\left(g_{h_1 h_2 h_2}\right)^2}{16\pi m_{h_1}}\left(1-\dfrac{4m_{h_2}^2}{m_{h_1}^2}\right)^{1/2}\,\,,
\end{equation}
and
\begin{equation}\label{eq:30}
\Gamma_1^{inv}\left(h_1\rightarrow h_3 h_3\right)=\dfrac{\left(g_{h_1 h_3 h_3}\right)^2}{16\pi m_{h_1}}\left(1-\dfrac{4m_{h_3}^2}{m_{h_1}^2}\right)^{1/2}\,\,,
\end{equation}
where the couplings $g_{h_1 h_2 h_2}$ and $g_{h_1 h_3 h_3}$ can be expressed in terms of the mixing angles, VEVs and the model parameters. We compute these two couplings numerically.
The invisible decay branching fraction for SM like scalar $h_1$ can be written as
\begin{equation}\label{eq:31}
{\rm{Br}^1}_{\text{inv}}=\dfrac{\Gamma_1^{\text{inv}}}{\Gamma_1}\,\,.
\end{equation}
We adopt the bounds on the signal strength of SM Higgs $R_1\ge0.84$ \cite{Khachatryan:2015, {Aad:2016}} and on the invisible decay branching fraction for SM Higgs, $\text{Br}_{\text{inv}}\leq 0.24$ \cite{CERN:2016} to further constrain the model parameters.

\section{Formation of Domain Walls and the consequent production of Gravitational Waves}

As mentioned earlier, a discrete symmetry $\mathcal{Z}_2 \times \mathcal{Z}^{\prime}_2$ is imposed on the potential in (Eq. \ref{eq:1}). When  $\mathcal{Z}_2$ and $\mathcal{Z}^{\prime}_2$ are spontaneously broken the fields $S_1$ and $S_2$ acquire non zero VEVs and domain walls are formed around the boundaries of the newly created domains. But these stable domain walls create problems in standard cosmology because if the energy density of the stable domain walls starts to dominate the energy density of the Universe, the Universe would have a rapid expansion ($\propto t^2$) which is incompatible with standard cosmology. Hence to avoid the domination of domain walls one can adopt a bias term \cite{Saikawa:2017hiv} to the scalar potential to destabilise the domain walls and eventually collapse them. This term creates an energy difference between the degenerate vacua and this energy difference affects the domain walls with a volume pressure. When this volume pressure exceeds the surface tension of the wall, annihilation of the domain walls takes place. During the process of annihilation of domain walls, some parts of its energy are converted to GWs and contribute as a stochastic background of GW in the present Universe.

\subsection{Domain Walls Formation}

In this work we consider planar domain walls perpendicular to the $z$ axis \cite{Hattori:2015xla} to be formed around the boundaries of the domains after the spontaneous breaking of the discrete symmetries $\mathcal{Z}$ and $\mathcal{Z}^\prime$. 
We define the scalars $S_1$ and $S_2$ as 
$S_1=\dfrac{v_1}{\sqrt{2}} e^{i \phi_1}$, $S_2=\dfrac{v_2}{\sqrt{2}} e^{i \phi_2}$
where we fix the radial part to their minima and introduce two phase factors $\phi_1$ and $\phi_2$. 
The potential in Eq. \ref{eq:1} can be rewritten as
\begin{equation}\label{eq:32}
\begin{aligned}
V(\phi_1, \phi_2)=\dfrac{1}{2}\mu_H^2 v^2 + \dfrac{1}{2} \mu_{S_1}^2 v_1^2 +\dfrac{1}{2} \mu_{S_2}^2 v_2^2 +\dfrac{1}{4} \lambda_H v^4 +\dfrac{1}{4} \lambda_{S_1} v_1^4 +\dfrac{1}{4} \lambda_{S_2} v_2^4 + \dfrac{1}{4} \lambda_{HS_1} v^2 v_1^2 + \dfrac{1}{4}\lambda_{HS_2} v^2 v_2^2 \\+\dfrac{1}{4} \lambda_{S_1 S_2} v_1^2 v_2^2+\dfrac{1}{4} g_{S_1 S_2} v_1^2 v_2^2 + \dfrac{1}{2} \kappa_{S_1S_2} v_1^2 v_2^2 \cos[2\left(\phi_2\left(z\right) -\phi_1\left(z\right)\right)]\,\,.\hspace{5cm}
\end{aligned}
\end{equation}
The kinetic part of the Lagrangian can be expressed in terms of $\phi_1$ and $\phi_2$ as
\begin{equation}\label{eq:33}
{\mathcal{L}_{\rm{kinetic}}}\left(\phi_1,\phi_2\right)= \dfrac{v_1^2}{2} \left(\partial_{\mu} \phi_1\right) \left(\partial^{\mu} \phi_1\right) + \dfrac{v_2^2}{2} \left(\partial_{\mu} \phi_2\right) \left(\partial^{\mu} \phi_2\right)\,\,.
\end{equation}
Using the potential in Eq. \ref{eq:32} the two field equations for $\phi_1$ and $\phi_2$
\begin{equation}\label{eq:34}
\partial_{\mu} \dfrac{{\mathcal{L}_{\rm{kinetic}}}}{\partial_{\mu}\left(\partial \phi_1 \right)}+\dfrac{\partial V}{\partial \phi_1}=0\,\,,
\end{equation}
\begin{equation}\label{eq:35}
\partial_{\mu} \dfrac{{\mathcal{L}_{\rm{kinetic}}}}{\partial_{\mu}\left(\partial \phi_2 \right)}+\dfrac{\partial V}{\partial \phi_2}=0\,\,,
\end{equation} 
can be obtained as  
\begin{equation}\label{eq:36}
\dfrac{d^2 \phi_1}{dz^2} - \dfrac{\kappa_{S_1 S_2} v_2^2}{2} \sin [2\left(\phi_1-\phi_2\right)]=0\,\,,
\end{equation}
\begin{equation}\label{eq:37}
\dfrac{d^2 \phi_2}{dz^2} + \dfrac{\kappa_{S_1 S_2} v_1^2}{2} \sin [2\left(\phi_1-\phi_2\right)]=0\,\,.
\end{equation}
In order to obtain the domain wall solution we solve the above two equations (Eqs. \ref{eq:36} and \ref{eq:37}) with the boundary conditions $\phi_i \rightarrow \dfrac{2\pi n}{2}$ at $z\rightarrow -\infty$ and $\phi_i \rightarrow \dfrac{2\pi \left(n+1\right)}{2}$ at $z\rightarrow \infty$ (with $i=1,2$ and $n$=0, 1, 2) \cite{Hattori:2015xla}. By considering $v_1^2=v_2^2$ (for analytical simplification) we get the following solution for $n=0$,
\begin{equation}\label{eq:38}
\phi_1=\tan^{-1} \left[\sqrt{2}\hspace{1mm} {\text{exp}}\left(2z\sqrt{\xi}\right)\right],
\end{equation}
where $\xi=\kappa_{S_1 S_2}\dfrac{v_2^2}{2}$. The solution of $\phi_2$ is obtained as $\phi_2=-\dfrac{v_1^2}{v_2^2} \phi_1$ i.e., $\phi_2=-\phi_1$ when $v_1^2$ and $v_2^2$ are equal. 
The width of the domain wall can be computed from the domain wall solution as
\begin{equation}\label{eq:39}
\delta\simeq\left(2 \sqrt{\xi}\right)^{-1}\,\,.
\end{equation}
The expression for the energy density $\rho_{\rm{wall}}$ of the domain walls is given by
\begin{equation}\label{eq:40}
\rho_{\rm{wall}}=\left(\left|\dfrac{dS_1 \left(z\right)}{dz}\right|^2 + \left|\dfrac{dS_2 \left(z\right)}{dz}\right|^2 + V\left(S_1\left(z\right),S_2\left(z\right),\dfrac{v}{\sqrt{2}} \right)-V\left(\dfrac{v_1}{\sqrt{2}},\dfrac{v_2}{\sqrt{2}},\dfrac{v}{\sqrt{2}} \right)\right)\,\,.
\end{equation}
Note that, we subtracted a constant term $V\left(\dfrac{v_1}{\sqrt{2}},\dfrac{v_2}{\sqrt{2}},\dfrac{v}{\sqrt{2}} \right)$ to satisfy the condition $\rho_{\rm{wall}}\rightarrow 0$ for $z\rightarrow \pm \infty $.
The domain wall tension $\sigma_{\rm{wall}}$ can be computed by integrating the energy density $\rho_{\rm{wall}}$ over the $z$ axis:
\begin{equation}\label{eq:41}
\sigma_{\rm{wall}}=\int dz \hspace{1mm}\rho_{\rm{wall}} \left(z\right)\,\,.
\end{equation}

\subsection{GWs Production from Domain Walls}

As discussed earlier, the unstable domain walls would collapse and produce a significant amount of GWs. In this section, we briefly present the calculations to obtain the energy densities of GWs produced from the annihilation of domain walls \cite{Hiramatsu:2013qaa, {Hiramatsu:2010yz}}.  

It is discussed before that domain walls must collapse or annihilate before they can overclose the Universe. The annihilation occurs when the volume pressure force tends to overcome the tension force. From this condition one can calculate the annihilation time using the expression \cite{Saikawa:2017hiv}
\begin{equation}\label{eq:43}
t_{\rm{ann}}=6.58\times 10^{-4} \hspace{1mm} {\rm{sec}}\hspace{1mm} C_{\rm{ann}} \hspace{1mm} \mathcal{A} \left(\dfrac{\sigma_{\rm{wall}}}{{\rm{TeV}}^3}\right) \left(\dfrac{V_{\rm{bias}}}{{\rm{MeV}}^4}\right)^{-1} \,\,,
\end{equation}
where $\mathcal{A}=0.8\pm0.1$ is the area parameter \cite{Caprini:2017vnn} and  $C_{\rm{ann}}$ is a model depended parameter of order $O(1)$. Here we take $C_{\rm{ann}}=5$ for $\mathcal{Z}_2$ symmetry \cite{Kawasaki:2014sqa}.  In the above equation (Eq. \ref{eq:43}), $V_{\rm bias}$ denotes the bias term of the potential introduced to make the domain walls unstable.  
The temperature at the annihilation time $t=t_{\rm{ann}}$ can be defined as \cite{Saikawa:2017hiv}
\begin{equation}\label{eq:44}
T_{\rm{ann}}=3.41\times 10^{-2} \hspace{1mm} {\rm{GeV}}\hspace{1mm} C_{\rm{ann}}^{-\frac{1}{2}} \hspace{1mm} \mathcal{A}^{-\frac{1}{2}} \left(\dfrac{g_{*}\left(T_{\rm{ann}}\right)}{10}\right)^{-\frac{1}{4}} \left(\dfrac{\sigma_{\rm{wall}}}{{\rm{TeV}}^3}\right)^{-\frac{1}{2}} \left(\dfrac{V_{\rm{bias}}}{{\rm{MeV}}^4}\right)^{\frac{1}{2}}\,\,,
\end{equation}
where $g_*(T_{\rm ann})$ is the relativistic degrees of freedom for the energy density at temeperatute $T_{\rm ann}$. From Eq. \ref{eq:43}, it is clear that $t_{\rm{ann}}$ depends on the bias term. Hence to annihilate the domain walls before they can overclose the Universe ($t_{\rm{ann}}<t_{\rm dominate}$), we require a lower bound on the magnitude of $V_{\rm{bias}}$ \cite{Saikawa:2017hiv}
\begin{equation}\label{eq:45}
V_{\rm{bias}}^{1/4}>2.18\times 10^{-5} \hspace{1mm} {\rm{GeV}}\hspace{1mm} C_{\rm{ann}}^{\frac{1}{4}} \hspace{1mm} \mathcal{A}^{\frac{1}{2}} \left(\dfrac{\sigma_{\rm{wall}}}{{\rm{TeV}}^3}\right)^{\frac{1}{2}}\,\,.
\end{equation}
This condition also constrains the value of the annihilation temperature $T_{\rm{ann}}$ \cite{Saikawa:2017hiv}
\begin{equation}\label{eq:46}
T_{\rm{ann}}>1.62\times 10^{-5} \hspace{1mm} {\rm{GeV}} \hspace{1mm} \mathcal{A}^{\frac{1}{2}} \left(\dfrac{g_{*}\left(T_{\rm{ann}}\right)}{10}\right)^{-\frac{1}{4}} \left(\dfrac{\sigma_{\rm{wall}}}{{\rm{TeV}}^3}\right)^{\frac{1}{2}}\,\,.
\end{equation}
If the domain walls decay into the SM particles, the decay products would affect the creation of light elements at the era of BBN, which is not favourable by the present observational results. Therefore it is required that the lifetime of domain walls should be shorter than $t_{\rm{ann}}\lesssim 0.1$ sec \cite{Reverberi:2014uva, {Santillan:2019oyy}}. This condition provides another bound on the magnitude of $V_{\rm{bias}}$ and this bound is given by \cite{Saikawa:2017hiv}
\begin{equation}\label{eq:47}
V_{\rm{bias}}^{1/4}>5.07\times 10^{-4} \hspace{1mm} {\rm{GeV}}\hspace{1mm} C_{\rm{ann}}^{\frac{1}{4}} \hspace{1mm} \mathcal{A}^{\frac{1}{4}} \left(\dfrac{\sigma_{\rm{wall}}}{{\rm{TeV}}^3}\right)^{\frac{1}{4}}\,\,.
\end{equation}
The peak intensity of GWs (produced from the annihilations of domain walls) at the present time $t_0$ is given by \cite{Saikawa:2017hiv}
\begin{equation}\label{eq:48}
\Omega^{\rm{dw}}_{\text{GW}}{\rm h}^2\left(t_0\right)_{{\rm{peak}}}=7.2\times 10^{-18} \hspace{1mm} \overline{\epsilon}_{{\rm{GW}}} \hspace{1mm} \mathcal{A}^{2} \left(\dfrac{g_{*s}\left(T_{\rm{ann}}\right)}{10}\right)^{-\frac{4}{3}} \left(\dfrac{\sigma_{\rm{wall}}}{{\rm{TeV}}^3}\right)^{2} \left(\dfrac{T_{\rm{ann}}}{10^{-2}{\rm{GeV}}}\right)^{-4}\,\,,
\end{equation}
where $\overline{\epsilon}_{{\rm{GW}}}=0.7\pm 0.4$ \cite{Caprini:2017vnn} is the efficiency parameter and $g_*(T_{\rm ann})$ is the relativistic degrees of freedom at annihilation temperature for the entropy density. It can be mentioned here that in our analysis we assume that the GWs are produced from the annihilation of domain walls in the radiation dominated era. 
The peak frequency of the GWs for this scenario can be estimated from the following expression \cite{Saikawa:2017hiv}
\begin{equation}\label{eq:49}
f^{\rm{dw}}\left(t_0\right)_{{\rm{peak}}}=1.1\times 10^{-9} \hspace{1mm}{\rm{Hz}} \hspace{1mm} \left(\dfrac{g_{*}\left(T_{\rm{ann}}\right)}{10}\right)^{\frac{1}{2}}  \left(\dfrac{g_{*s}\left(T_{\rm{ann}}\right)}{10}\right)^{-\frac{1}{3}} \left(\dfrac{T_{\rm{ann}}}{10^{-2}{\rm{GeV}}}\right)\,\,.
\end{equation}
In Ref. \cite{Gelmini:1988sf}, the authors mentioned that the domain walls cannot be formed if the bias term $V_{\rm{bias}}$ is large enough. Hence an upper bound  $V_{\rm{bias}}<< V$ should also be considered. 
We take $V_{\rm{bias}}$ as the free parameter of the model. By varying $V_{\rm{bias}}$ and by calculating  model-dependent domain wall tension we compute the GW intensities. Needless to mention that when we choose $V_{\rm{bias}}$ the constraints on $V_{\rm{bias}}$ expressed in Eqs. \ref{eq:45}, \ref{eq:47} are respected.

There are some recent studies on GW from domain walls based on numerical simulations that show $\Omega^{\rm{dw}}_{\text{GW}}{\rm h}^2 \propto f^3$ for $f< f^{\rm{dw}}_{{\rm{peak}}}$ and $\Omega^{\rm{dw}}_{\text{GW}}{\rm h}^2 \propto \dfrac{1}{f}$ for $f\geqslant f^{\rm{dw}}_{{\rm{peak}}}$ \cite{Hiramatsu:2013qaa}.
In this work we use these relations and choose five sets of BPs (related to our model parameters) to calculate the GW intensity and peak frequency in the case of the present particle physics model.
\section{Electroweak Phase transition and Gravitational Waves Production from Strong First-Order Phase Transition}
In this section, we pursue the electroweak phase transition and possible production mechanism of GWs from SFOPT in our proposed two complex scalars extended SM.
\subsection{Finite Temperature Effective Potential}
In order to explore the electroweak phase transition (EWPT) in the present model we include two potential terms with the tree-level potential $V_{\text{}tree-level}$. Now the modified finite temperature effective potential can be expressed as \cite{Wainwright:2012}
\begin{equation}\label{eq:50}
V_{\text{eff}}=V_{\text{}tree-level}+V_{1-loop}^{T=0}+V_{1-loop}^{T\neq0}\,\,,
\end{equation}
where $V_{1-loop}^{T=0}$ is the zero temperature Coleman-Weinberg one-loop effective potential and $V_{1-loop}^{T\neq0}$ is the finite
temperature one-loop effective potential. The zero-temperature tree-level potential $V_{\rm tree-level}^{T=0}$ can be expressed in terms of $v$, $v_1$, $v_2$ as  
\begin{equation}\label{eq:51}
\begin{aligned}
V_{\rm tree-level}^{T=0}=\dfrac{1}{2}\mu_H^2 v^2 + \dfrac{1}{2} \mu_{S_1}^2 v_1^2 +\dfrac{1}{2} \mu_{S_2}^2 v_2^2 +\dfrac{1}{4} \lambda_H v^4 +\dfrac{1}{4} \lambda_{S_1} v_1^4 +\dfrac{1}{4} \lambda_{S_2} v_2^4 + \dfrac{1}{4} \lambda_{HS_1} v^2 v_1^2 + \dfrac{1}{4}\lambda_{HS_2} v^2 v_2^2 \\+\dfrac{1}{4} \lambda_{S_1 S_2} v_1^2 v_2^2+\dfrac{1}{4} g_{S_1 S_2} v_1^2 v_2^2 + \dfrac{1}{4} \kappa_{S_1S_2} v_1^2 v_2^2\,\,.\hspace{7cm}
\end{aligned}
\end{equation}
We obtain the above equation (Eq. \ref{eq:51}) from Eq. \ref{eq:1} by replacing the scalar fields $H$, $S_1$, $S_2$ with their VEVs $v$, $v_1$, $v_2$ respectively. The zero-temperature Coleman-Weinberg one-loop effective potential can be written as \cite{Wainwright:2012}
\begin{equation}\label{eq:52}
V_{1-loop}^{T=0}=\pm \dfrac{1}{64\pi^2} \sum_{i} n_i m_i^4 \left[ \log\dfrac{m_i^2}{Q^2}-C_i\right]\,\,,
\end{equation}
where the `+' and `$-$' symbols refer to the sign of the bosons and fermions. The summation $i$ is over all the particles associated in the model and $i\equiv(h_1,h_2,h_3,G^{\pm},G^{0},\chi_1,\chi_2,W,Z,t$). In Eq. \ref{eq:52} $n_i$, $m_i$ and $C_i$ denote the number of degrees of freedom, the field-dependent masses and renormalisation-scheme-dependent numerical constant of the particle $i$ respectively. The degrees of freedom of the above mentioned particle species are $(n_{W^{\pm}})_L=4$, $(n_{W^{\pm}})_T=2$, $(n_Z)_L=2$, $(n_Z)_T=1$, $n_t=12$, $n_{G^{\pm}}=2$ and $n_{h_,h_2,h_3,G^0,\chi_1,\chi_2}=1$. The numerical values of the constant $C_i$ are $(C_{W,Z})_T=1/2$ for the transverse component of $W, \hspace{1mm} Z$ boson, $(C_{W,Z})_L=3/2$ for the longitudinal component of $W, \hspace{1mm} Z$ boson and for the other particle species, $C_{h_,h_2,h_3,G^+,G^-,G^0,\chi_1,\chi_2,t}=3/2$. The field-dependent masses of the gauge bosons $m_W$, $m_Z$ and the top quark $m_t$ in terms of $v$, $v_1$, $v_2$ at $T=0$ can be expressed as

\begin{equation}\label{eq:54}
m_W^2=\dfrac{1}{4}g^2 v^2\,\,,
\end{equation}

\begin{equation}\label{eq:55}
m_Z^2=\dfrac{1}{4}\left(g^2+{g^{\prime}}^2\right) v^2\,\,,
\end{equation}

\begin{equation}\label{eq:53}
m_t^2=\dfrac{1}{2}y_t^2 v^2 \,\,,
\end{equation}
where $g$, $g^{\prime}$ and $y_t$ refer to the $SU(2)_L$ gauge coupling, $U(1)_Y$ gauge coupling and top Yukawa coupling of the SM respectively. We perform our analysis by considering the Landau gauge where the Goldstone bosons are massless at zero temperature theory and also consider that ghost contributions are not appearing here  \cite{Basler:2016obg}.
In Eq. \ref{eq:52}, the quantity $Q$ is the renormalisation scale which we fix as  $Q=v=246.22$ GeV in our calculations. The finite temperature one-loop effective potential is given by \cite{Wainwright:2012}
\begin{equation}\label{eq:56}
V_{1-loop}^{T\neq0}=\dfrac{T^4}{2\pi^2} \sum_{i} n_i J_{\pm}\left[\dfrac{m_i^2}{T^2}\right]\,\,,
\end{equation}
with
\begin{equation}\label{eq:57}
J_{\pm}\left(\dfrac{m_i^2}{T^2}\right)=\pm \int_0^{\infty} dy \hspace{1mm} y^2 \log\left(1\mp e^{-\sqrt{y^2+\dfrac{m_i^2}{T^2}}}\right)\,\,.
\end{equation}
Here we would like to mention that the values of the classical VEVs i.e ($v,\hspace{1mm}v_1,\hspace{1mm}v_2$) change with temperature and at $T=0$ it tends to the classical fixed values.
We apply daisy resummation method \cite{Arnold} in our work for executing the thermal correction to the boson masses as 
$\mu^{\prime 2}_{H}(T)=\mu_H^2 + c_1 T^2$\,\,, 
$\mu^{\prime 2}_{S_1}(T)=\mu_{S_1}^2+c_2 T^2$ and 
$\mu^{\prime 2}_{S_2} (T)=\mu_{S_2}^2+c_3 T^2$\,\,, where
\begin{equation}\label{eq:58}
c_1=\dfrac{6\lambda_H+2\lambda_{HS_1}+2\lambda_{HS_2}}{12}+\dfrac{3g^2+{g^{\prime}}^2}{16}+\dfrac{y_t^2}{4}\,\,,
\end{equation}

\begin{equation}\label{eq:59}
c_2=\dfrac{6\lambda_{S_1}+2\lambda_{HS_1}+2\lambda_{S_1 S_2}+g_{S_1 S_2}}{12}\,\,,
\end{equation}
and
\begin{equation}\label{eq:60}
c_3=\dfrac{6\lambda_{S_2}+2\lambda_{HS_2}+2\lambda_{S_1 S_2}+g_{S_1 S_2}}{12}\,\,.
\end{equation}
For the gauge bosons there are only finite temperature corrections to the longitudinal components and the corrected thermal masses of $W$ and $Z$ boson are given by \cite{Basler:2016obg}
\begin{equation}\label{eq:61}
m_W^2\left(T\right)=m_W^2 +2g^2T^2\,\,,
\end{equation}
and
\begin{equation}\label{eq:62}
m_Z^2\left(T\right)=\dfrac{1}{2}m_Z^2 +\left(g^2+{g^{\prime}}^2\right)T^2 + \dfrac{1}{8}\sqrt{\left[\left(g^2-{g^{\prime}}^2\right)^2 \left(64 T^2+16 T^2 v^2\right)+\left(g^2+{g^{\prime}}^2\right)^2v^4\right]}\,\,.
\end{equation}
We use the publicly available package CosmoTransitions \cite{Wainwright:2012} to include the finite temperature corrections with the tree-level potential.

\subsection{Gravitational Waves Production from Strong First-Order Phase Transition}

In this section, we discuss how the gravitational waves are produced from the SFOPT principle. Initially, the Universe is in a global minimum and with the decrease in temperature as the Universe evolves a minimum develops (metastable state). When the newly formed minima as a function of temperature tends to the true minimum then the process of tunnelling from false (global) to true minimum takes place through the nucleation of bubbles. During the process, a situation comes when the two minima become degenerate and that temperature indicates the critical temperature $T_c$. At the nucleation temperature (less than $T_c$) the phase transition completes where at least one bubble is formed per unit time per Hubble volume. The latent heat energy liberated during the phase of first-order phase transition process contributes to the energy density of the stochastic GW background. The production of GWs is principally caused by three mechanisms namely (i) bubble collisions, (ii) sound waves induced by the bubbles running through the cosmic plasma and (iii) turbulence induced by the bubble expansions in the cosmic plasma. 

The probability of formation of a bubble per unit time per unit volume at a temperature $T$ can be written as \cite{Linde:1983}
\begin{equation}\label{eq:63}
\Gamma=\Gamma_0\left(T \right) e^{-S_3\left(T \right)/T}\,\,,
\end{equation}
where $\Gamma_0\left(T \right)$ scales as $\Gamma_0\left(T \right)\propto T^4$ and the Euclidean action of the critical bubble $S_3\left(T \right)$ is given by \cite{Linde:1983}
\begin{equation}\label{eq:64}
S_{3}=4\pi \int dr \hspace{1mm}r^{2} \left[ \dfrac{1}{2} \left(\partial_{r} \vec{\phi} \right)^2 +V_{eff}\right]\,\,,
\end{equation}
where $\vec{\phi}=(h,s_1,s_2)$ represents a vector of the scalar fields in the potential $V$ (Eq. \ref {eq:1}) and $V_{eff}$ is the finite temperature effective potential expressed in Eq. \ref{eq:50}. Nucleation of bubbles occurs at a temperature $T_n$ where $T_n$ is the temperature where the conditions $\Gamma\sim 1$ and $S_3\left(T_n \right)/T_n\approx 140$ \cite{Wainwright:2012} are obeyed.

As mentioned, the production mechanism of GWs from the strong first-order electroweak phase transition are driven by three processes namely bubble collisions \cite{Kosowsky:1992a}-\cite{Caprini:2008}, sound waves induced by the bubbles running through the cosmic plasma \cite{Hindmarsh:2014}-\cite{Hindmarsh:2015} and turbulence induced by the bubble expansions in the cosmic plasma \cite{Caprini:2006}-\cite{Caprini:2009}. The total GW intensity $\Omega_{\text{GW}}{\rm h}^2$ from the SFOPT for a particular frequency $f$ can be obtained by adding the contributions of the above three mechanisms and we have \cite{Kosowsky:1992a}-\cite{Hindmarsh:2020hop}
\begin{equation}\label{eq:65}
\Omega^{\rm{PT}}_{\text{GW}}{\rm h}^2=\Omega_{\text{col}}{\rm h}^2+ \Omega_{\text{SW}}{\rm h}^2+ \Omega_{\text{turb}}{\rm h}^2\,\,.
\end{equation}
The contribution of GW intensity from the bubbles collision $\Omega_{\text{col}}{\rm h}^2$ is given by
\begin{equation}\label{eq:66}
\Omega_{\text{col}}{\rm h}^2=1.67\times 10^{-5} \left(\dfrac{\beta}{H} \right) ^{-2} \dfrac{0.11 v_{w}^3}{0.42+v_{w}^2} \left(\dfrac{\kappa \alpha}{1+\alpha}\right)^2 \left(\dfrac{g_*}{100}\right)^{-\frac{1}{3}}\dfrac{3.8 \left(\dfrac{f}{f_{col}}\right)^{2.8}}{1+2.8 \left(\dfrac{f}{f_{\text{col}}}\right)^{3.8}}\,\,\,,
\end{equation}
where $\beta$ is the inverse time scale of the phase transition parameter and it has the following form 
\begin{equation}\label{eq:67}
\beta=\left[H T \dfrac{d}{dT}\left( \dfrac{S_3}{T}\right) \right]\bigg|_{T_n}\,\,,
\end{equation}
where $H$ denotes the Hubble parameter at the nucleation temperature $T_n$. We estimate the bubble wall velocity $v_w$ using the following expression as \cite{Steinhardt:1982, {Kamionkowski:1994}, {Dev:2019njv}, {Paul:2019pgt}}
\begin{equation}\label{eq:68}
v_w=\dfrac{1/\sqrt{3}+\sqrt{\alpha^2+2\alpha/3}}{1+\alpha}\,\,.
\end{equation}
In Eq. \ref{eq:66} the quantity $\kappa$ is the fraction of latent heat energy deposited in a thin shell which takes the form
\begin{equation}\label{eq:69}
\kappa=1-\dfrac{\alpha_{\infty}}{\alpha}\,\,,
\end{equation}
with \cite{Shajiee:2018jdq,{Caprini:2015zlo}}
\begin{equation}\label{eq:70}
\alpha_{\infty}=\dfrac{30}{24\pi^2 g_{*}} \left(\dfrac{v_n}{T_n} \right)^2 \left[6 \left( \dfrac{m_W}{v}\right)^2 +3\left( \dfrac{m_Z}{v}\right)^2 +6\left( \dfrac{m_{t}}{v}\right)^2\right]\,\,.
\end{equation}
where $v_n$ is the VEV of Higgs at the nucleation temperature $T_n$, $m_W$, $m_Z$ and $m_t$ denotes the masses of the gauge bosons $W$, $Z$ and the top quark $t$ respectively. The parameter $\alpha$ can be defined as the ratio of energy density difference between false and true vacuum released during the electroweak phase transition $\rho_{\text{vac}}$ to the background energy density of the plasma $\rho_*^{\text{rad}}$ at $T_n$. The expression of $\alpha$ is given by
\begin{equation}\label{eq:71}
\alpha=\left[\dfrac{\rho_{\text{vac}}}{\rho^*_{\text{rad}}}\right]\bigg|_{T_n}\,\,.
\end{equation}
with
\begin{equation}\label{eq:72}
\rho_{\text{vac}}=\left[\left(V_{\text{eff}}^{\text{high}}-T\dfrac{dV_{\text{eff}}^{\text{high}}}{dT} \right)-\left(V_{\text{eff}}^{\text{low}}-T\dfrac{dV_{\text{eff}}^{\text{low}}}{dT} \right)\right]\,\,,
\end{equation}
and
\begin{equation}\label{eq:73}
\rho^*_{\text{rad}}=\dfrac{g_* \pi^2 T_n^4}{30}\,\,.
\end{equation}
The expression of peak frequency $f_\text{col}$ (Eq. \ref{eq:66}) produced by the bubble collisions is 
\begin{equation}\label{eq:74}
f_{\text{col}}=16.5\times10^{-6}\hspace{1mm} \text{Hz} \left( \dfrac{0.62}{v_{w}^2-0.1 v_w+1.8}\right)\left(\dfrac{\beta}{H} \right) \left(\dfrac{T_n}{100 \hspace{1mm} \text{GeV}} \right) \left(\dfrac{g_*}{100}\right)^{\frac{1}{6}}\,\,.
\end{equation}
The contribution of GW intensity from the sound wave (SW) (Eq. \ref{eq:65}) is given by
\begin{equation}\label{eq:75}
\Omega_{\text{SW}}{\rm h}^2=2.65\times 10^{-6} \left(\dfrac{\beta}{H} \right) ^{-1} v_{w} \left(\dfrac{\kappa_{v} \alpha}{1+\alpha}\right)^2 \left(\dfrac{g_*}{100}\right)^{-\frac{1}{3}}\left(\dfrac{f}{f_{\text{SW}}}\right)^{3} \left[\dfrac{7}{4+3 \left(\dfrac{f}{f_{\text{SW}}}\right)^{2}}\right]^{\frac{7}{2}}\,\,,
\end{equation}
where $\kappa_v$ is the fraction of latent heat energy converted into the bulk motion of the fluid which takes the form
\begin{equation}\label{eq:76}
\kappa_v=\dfrac{\alpha_{\infty}}{\alpha}\left[ \dfrac{\alpha_{\infty}}{0.73+0.083\sqrt{\alpha_{\infty}}+\alpha_{\infty}}\right]\,\,.
\end{equation}
The peak frequency $f_{\text{SW}}$ for the sound wave contribution is
\begin{equation}\label{eq:77}
f_{\text{SW}}=1.9\times10^{-5}\hspace{1mm} \text{Hz} \left( \dfrac{1}{v_{w}}\right)\left(\dfrac{\beta}{H} \right) \left(\dfrac{T_n}{100 \hspace{1mm} \text{GeV}} \right) \left(\dfrac{g_*}{100}\right)^{\frac{1}{6}}\,\,.
\end{equation}
The authors in Refs. \cite{Caprini:2015zlo,Ellis:2018mja,Ellis:2019oqb} proposed that the contribution of SW to the total GW intensity depends on the Hubble time scale. If it survives more than a Hubble time then the expression in  Eq. \ref{eq:75} will be valid otherwise it is an overestimation to the GW signal. Therefore from the following Refs. \cite{Caprini:2015zlo,Ellis:2018mja,Ellis:2019oqb} we estimate a factor $\dfrac{HR_*}{\bar{U}_f}$ $\big({HR_*}/{\bar{U}_f}$ called the suppression factor, where $\bar{U}_f$ is the root-mean-square (RMS) fluid velocity and $R_*$ is the mean bubble separation$\big)$ to verify whether the SW components survive more than a Hubble time. If the suppression factor turns out to be less than 1, then  we need to include the factor to the sound wave component of the GW intensity.

The contribution $\Omega_{\text{turb}}{\rm h}^2$ from the turbulence in the plasma is given by
\begin{equation}\label{eq:78}
\Omega_{\text{turb}}{\rm h}^2=3.35\times 10^{-4} \left(\dfrac{\beta}{H} \right) ^{-1} v_{w} \left(\dfrac{\epsilon \kappa_v \alpha}{1+\alpha}\right)^{\frac{3}{2}} \left(\dfrac{g_*}{100}\right)^{-\frac{1}{3}} \dfrac{\left(\dfrac{f}{f_{\text{turb}}}\right)^{3}\left( 1+\dfrac{f}{f_{\text{turb}}}\right)^{-\frac{11}{3}}}{\left(1+\dfrac{8\pi f}{h_{*}}\right)}\,\,,
\end{equation}
where $\epsilon=0.1$ and the peak frequency $f_{\text{turb}}$ for the turbulence mechanism reads:
\begin{equation}\label{eq:79}
f_{\text{turb}}=2.7\times10^{-5}\hspace{1mm} \text{Hz} \left( \dfrac{1}{v_{w}}\right)\left(\dfrac{\beta}{H} \right) \left(\dfrac{T_n}{100 \hspace{1mm} \text{GeV}} \right) \left(\dfrac{g_*}{100}\right)^{\frac{1}{6}}\,\,,
\end{equation}
where the parameter
\begin{equation}\label{eq:80}
h_{*}=16.5\times10^{-6}\hspace{1mm} \text{Hz} \left(\dfrac{T_n}{100 \hspace{1mm} \text{GeV}} \right) \left(\dfrac{g_*}{100}\right)^{\frac{1}{6}}\,\,.
\end{equation}
We use Eqs. \ref{eq:65} - \ref{eq:80} for the estimation of the gravitational wave intensity in this work.

\section{Calculations and Results}

In this section, we estimate the intensities of the gravitational wave produced via domain wall annihilation as well as from strong first-order phase transition for the present extended SM. To check its detectability, our calculated model-dependent GW intensities are compared with the future space-based and ground-based interferometers such as ALIA, BBO, DECIGO, TianQin, Taiji, aLIGO, aLIGO+ and pulsar timing arrays such as SKA, IPTA, EPTA, PPTA, NANOGrav11 and NANOGrav12.5. For the calculation of GW intensities we choose five benchmark points (BPs) such that the numerical values of the model parameters in each of the chosen BPs satisfy the theoretical constraints (such as vacuum stability, perturbativity) as well as the experimental constraints (such as LHC). In case of GW production from domain walls, GW intensity depends mainly on factors such as domain wall tension $\sigma_{\rm{wall}}$, annihilation time $t_{\rm{ann}}$, annihilation temperature $T_{\rm{ann}}$, degrees of freedom at annihilation time $g_{*s}\left(T_{\rm{ann}}\right)$ and bias term $V_{\rm{bias}}$. On the other hand for the GW production via SFOPT, the strength of the first-order phase transition parameter $\alpha$, inverse of the time-scale of the phase transition parameter $\beta$, bubble wall velocity $v_{w}$, nucleation temperature $T_n$ and value of Higgs VEV $v_n$ at $T_n$ play significant roles. Using the equations given in section 4, we compute and explore the formation of domain walls and production of GW from the annihilation of the domain walls for the present model considered in this work. In order to calculate the domain wall tension we first solve the Eqs. \ref{eq:36} and \ref{eq:37} and get the domain wall solution. Then we estimate the domain wall tension using Eq. \ref{eq:40} and \ref{eq:41}. The domain walls are then made unstable by introducing a bias term $V_{\rm{bias}}$. Here we take $V_{\rm{bias}}$ as a free parameter. We also use two bounds on $V_{\rm{bias}}$ to make the domain walls annihilate before it could overclose the Universe and before BBN era. Considering these bounds we compute the annihilation time of the domain walls and annihilation temperature at that time by using Eq. \ref{eq:43} and Eq. \ref{eq:44} respectively. 
Finally, we calculate the GW intensity and the peak frequency at the present time from the annihilation of domain walls using Eqs. \ref{eq:48} and \ref{eq:49} respectively. Note that, our chosen five BPs for the calculation of GW intensity satisfy the bounds on $V_{\rm{bias}}$ and $T_{\rm{ann}}$.

\begin{table}[H]
\centering
\footnotesize
\begin{tabular}{|l|c|c|c|c|c|c|c|c|c|r|}
\hline
BP&$m_{h_{2}}$&$m_{h_{3}}$&$\theta_{12}$&$\theta_{13}$&$\theta_{23}$&$v_{1}$&$v_{2}$&$g_{S_1 S_2}$&$\kappa_{S_1 S_2}$\\
 &in GeV&in GeV&&&&in GeV&in GeV&&\\
 \hline
1&100&100&0.4&0.1&-0.9&200&200&0.03&-10\\ 
 \hline
2&150&50&0.1&0.05&0.01&250&250&0.1&-12.11\\ 
 \hline
3&200&200&-0.18&0.1&1.1&400&400&0.3&-10\\
\hline
4&500&500&-0.18&0.1&1.1&5000&5000&0.4&-10\\
\hline
5&500&400&0.4&0.1&-0.9&1000&1000&0.3&-10\\
\hline
\end{tabular}
\caption{The chosen five benchmarks points (BPs, BP 1-5) to explore the GW production from both the domain walls and strong first-order phase transition in two complex scalars extended SM.}\label{t1}
\end{table}

On the other hand, for the calculation of GW intensity from the first-order phase transition we consider a finite temperature effective potential (discussed in section 5). The finite temperature effective potential is obtained by adding two terms with the tree-level potential - one is the one-loop CW potential at zero temperature $V_{1-loop}^{T=0}$ and the other is the finite temperature potential $V_{1-loop}^{T\neq0}$. We consider the three possible processes such as bubble collisions, sound waves and turbulence in the plasma for the production of GWs from phase transition. Then we compute the total GW intensity using Eqs. \ref{eq:65} - \ref{eq:80} from the SFOPT. We also mention that in this work we choose five BPs to demonstrate that our model can induce strong first-order phase transition for each of the five choices which is a necessary condition for the production of GWs \cite{Basler:2016obg, {Moore:1998swa}}. The choice of the five BPs with the values of the model parameters are furnished in Table \ref{t1}. Here we use a package namely Cosmotransition for calculating the thermal parameters related to the phase transition properties for all the chosen BPs. 

We tabulate our calculated results in Tables \ref{t2}-\ref{t5}. While the calculations related to the GWs from domain walls are furnished in Table \ref{t2} and Table \ref{t4}, in Table \ref{t3} and Table \ref{t5} we show the computed results when GWs are formed via SFOPT. In Table \ref{t2} we show our results for the quantities $g_{*}\left(T_{\rm{ann}}\right)$, $g_{*s}\left(T_{\rm{ann}}\right)$, $t_{\rm{ann}}$, $T_{\rm{ann}}$, $V_{\rm{bias}}$, $ \sigma_{\rm{wall}}$ (related to the estimation of GW intensity from domain walls) for the five chosen BPs. We present our calculated GW intensity at the present epoch and the corresponding peak frequency for all the BPs (BP 1-5) in Table \ref{t4}. 

The values of the thermal parameters ($v_c$, $T_c$, $v_n$, $T_{n}$, $\alpha$, ${\beta}/{H}$) which are the useful parameters for the calculation of GW intensity from SFOPT are tabulated in Table \ref{t3} for all the five BPs. In Table \ref{t3} we also indicate the critical temperature $T_c$ (the temperature at which two phases are degenerate) and the numerical value of the $v_c$, the VEV at that temperature $T_c$. Note that all the BPs satisfy the condition $v_c/T_c\geqslant 1$ except the ones in BP 4. For BP 4 $v_c/T_c <  1$, so in that case the SFOPT condition is not satisfied and hence it is unable to produce GWs from SFOPT. So in further calculations of GW intensity we do not calculate the intensity for BP 4. Table \ref{t3} shows that the nucleation temperature $T_n$ is less than $T_c$. Therefore with further decrease in temperature from $T_c$, nucleation of the bubble occurs and phase transition is completed. In Table \ref{t3}, we also present the value of $\dfrac{HR_*}{\bar{U}_f}$, the factor which indicates whether or not the contribution of sound wave is survives more than Hubble time. In case $\dfrac{HR_*}{\bar{U}_f}< 1$, the SW contribution decays within the Hubble time. Following Refs. \cite{Caprini:2015zlo,Ellis:2018mja,Ellis:2019oqb} we compute this factor for all BPs and obtain $\dfrac{HR_*}{\bar{U}_f}< 1$ for all the BPs (BP 1, BP 2, BP 3, BP 5) except BP 4. Therefore as mentioned in Section 5 we multiply the factor with the sound wave component of GW intensity (Eq. \ref{eq:75}). 
In Table \ref{t5} we show the results for GW intensity and corresponding peak frequency from SFOPT. We obtain a single peak in GW intensity for the case of each BP but when BP 3 is adopted, two peaks appears instead for GW intensity when varied with GW frequency. The reason behind the two peaks being two GW components, bubble collisions and SW, play dominant roles. While in the other cases only bubble collision plays a major role and only bubble collision dominantly influences the peak of the GW spectrum. In case of BP 1, 2 and 5 the peak frequencies of the total GW spectrum from SFOPT are equal to the peak frequencies of the component of bubble collision $f_{\rm{col}}=f^{PT}_{\rm{peak}}$. However in the case of BP 3 for which two peaks are obtained, one peak corresponds to $f_{\rm{col}}$ and while the other is at $f_{\rm{SW}}$. Note that, since BP 4 does not satisfy the SFOPT criteria, it can not produce GWs from strong first-order phase transition. 

\begin{table}[H]
\centering
\small
\begin{tablenotes}
\centering
\item $C_{\rm{ann}}$=5, $\mathcal{A}$=0.8, $\overline{\epsilon}_{{\rm{GW}}}$=0.7.
\end{tablenotes}
\begin{tabular}{|l|c|c|c|c|c|c|c|c|c|r|}
\hline
BP & $g_{*}\left(T_{\rm{ann}}\right)$ & $g_{*s}\left(T_{\rm{ann}}\right)$ & $t_{\rm{ann}}$ & $T_{\rm{ann}}$ & $V_{\rm{bias}}$ & $ \sigma_{\rm{wall}}$\\
 & & &in sec& in GeV & in ${\rm{MeV}}^4$ &in ${\rm{TeV}}^3$\\
 \hline
1 & 10.57 & 10.57 & 0.0094 & 0.0089 & 0.01 & 3.58$\times10^{-2}$\\
\hline
2 & 10.75 & 10.75 & 0.0067 & 0.010 & 0.03 & 7.69$\times10^{-2}$\\
\hline
3 & 10.57 & 10.57 & 0.0094 & 0.0089 & 0.08 & 0.286\\
\hline
4 & 10.57 & 10.57 & 0.0099 & 0.0087 & 148 & 559\\
\hline
5 & 10.57 & 10.57 & 0.0098 & 0.0087 & 1.2 & 4.47\\
\hline
\end{tabular}
\caption{The values of the parameters used for the estimation of the GW intensity from the annihilation of domain walls for the chosen five BPs (BP 1-5). See text for details.}\label{t2}
\end{table}

\begin{table}[H]
\centering
\begin{tabular}{|l|c|c|c|c|c|c|c|c|c|r|}
\hline
BP & $f^{\rm{dw}}\left(t_0\right)_{{\rm{peak}}}$&$\Omega^{\rm{dw}}_{\text{GW}}{\rm h}^2\left(t_0\right)_{{\rm{peak}}}$\\
 & in Hz &\\
 \hline
1 & 9.87$\times10^{-10}$ & 6.14$\times10^{-21}$\\
\hline
2 & 1.16$\times10^{-9}$ & 1.45$\times10^{-20}$\\
\hline
3 & 9.87$\times10^{-10}$ & 3.93$\times10^{-19}$\\
\hline
4 & 9.61$\times10^{-10}$ & 1.67$\times10^{-12}$\\
\hline
5 & 9.67$\times10^{-10}$ & 1.04$\times10^{-16}$\\
\hline
\end{tabular}
\caption{The calculated values of peak frequency and its corresponding GW intensity from the annihilations of domain walls for the chosen five BPs (BP 1-5).}\label{t4}
\end{table}

\begin{table}[H]
\centering
\footnotesize
\begin{tabular}{|l|c|c|c|c|c|c|c|c|c|r|}
\hline
BP & $v_c$ & $T_c$ & $\dfrac{v_c}{T_c}$ & $v_n$ & $T_{n}$ & $\alpha$ & $\dfrac{\beta}{H}$ &
$\dfrac{HR_*}{\bar{U}_f}$\\
 & in GeV & in GeV && in GeV &in GeV&&&\\
 \hline
1 & 180.29 & 129.52 & 1.39 & 212.39 & 107.47 & 0.37 & 1042.83 & 0.62\\
\hline
2 & 140.42 & 135 & 1.04 & 190.98 & 99.80 & 0.33 & 819.36 & 0.82\\
\hline
3 & 167.63 & 95.57 & 1.75 & 183.69 & 90.43 & 0.029 & 2206.93 & 0.06\\
\hline
4 & 109.85 & 323.15 & 0.34 & 118.11 & 307.05 & 0.16 & 4473.43 & 11.43\\
\hline
5 & 166.10 & 117.04 & 1.42 & 174.89 & 110.04 & 0.44 & 2644.45 & 0.51\\
\hline
\end{tabular}
\caption{The values of the parameters related to the phase transition properties used for the estimation of the GW intensity from strong first-order phase transition for the chosen five BPs (BP 1-5). See the text for details.}\label{t3}
\end{table}

\begin{table}[H]
\centering
\begin{tabular}{|l|c|c|c|c|c|c|c|c|c|r|}
\hline
BP & $f^{\rm{PT}}_{{\rm{peak}}}$&$\Omega^{\rm{PT}}_{\text{GW}}{\rm h}^2_{{\rm{peak}}}$\\
 & in Hz &\\
 \hline
1 & 4.7$\times10^{-3}$ & 6.21$\times10^{-14}$\\
\hline
2 & 3.4$\times10^{-3}$ & 8.32$\times10^{-14}$\\
\hline
3 & 9.3$\times10^{-3}$, 5.5$\times10^{-2}$ & 1.4$\times10^{-17}$, 1.42$\times10^{-17}$\\
\hline
5 & 1.2$\times10^{-2}$ & 1.29$\times10^{-14}$\\
\hline
\end{tabular}
\caption{The calculated values of peak frequency and its corresponding GW intensity from the contribution of strong first-order phase transition for the chosen BPs (BP 1, BP 2, BP 3 and BP 5). We obtain two peak frequencies for BP 3 and its corresponding GW intensities are also mentioned in Table \ref {t5}. Note that we do not show the results for
BP 4 since in the case of BP 4 the strong first-order phase transition condition is not satisfied ($v_c/T_c<1$, Table \ref {t3}), as a result it is unable to produce GWs.}\label{t5}
\end{table}

\begin{figure}[H]
\centering
\includegraphics[width=8cm,height=6cm]{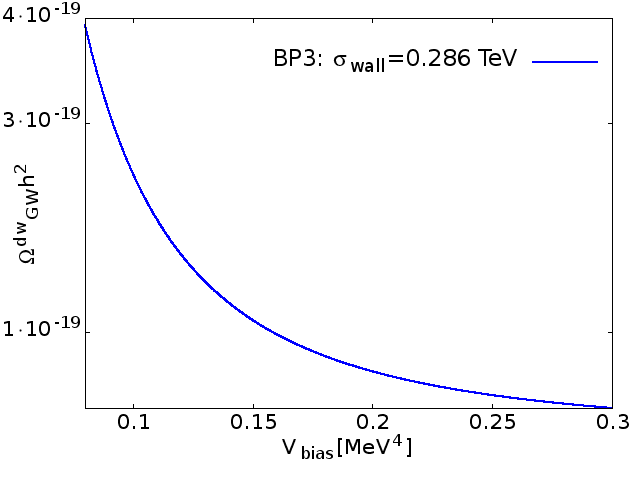}
\caption{The variation of the GW intensity $\Omega^{\rm{dw}}_{\text{GW}}{\rm h}^2$ from domain walls with biased potential $V_{\rm{bias}}$ for BP 3.}
\label{fig:1}
\end{figure}

\begin{figure}[H]
\centering
\includegraphics[width=8cm,height=6cm]{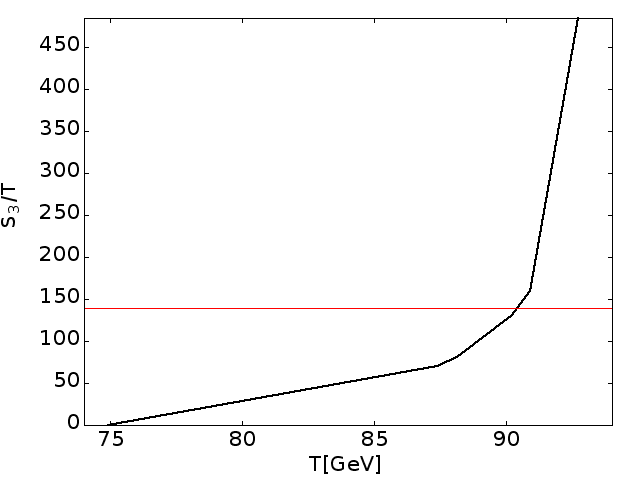}
\caption{Variation of the parameter $S_3 /T$ as a function of temperature for BP 3. The red horizontal line shows the condition $S_3 /T =140$, where the bubble nucleation occurs.}
\label{fig:2}
\end{figure}

In Figure \ref{fig:1}, we show the variation of GW intensity from domain walls as a function of $V_{\rm{bias}}$ with a fixed value for domain wall tension $\sigma_{\rm{wall}}$. In Figure \ref{fig:1} we adopt the benchmark values in BP 3 to demonstrate the variation of GW intensity with bias potential $V_{\rm{bias}}$. From Figure \ref{fig:1} one observes that smaller the value of $V_{\rm{bias}}$, more is the strength of GW intensity. That smaller values of $V_{\rm bias}$ lead to larger intensity of GW can also be concluded from Eqs. \ref{eq:44} and \ref{eq:48}. We repeat the computation of the variations of the GW intensity with $V_{\rm bias}$ for all the other BPs and obtain similar results. Therefore we choose small values of $V_{\rm{bias}}$ which satisfy the bounds mentioned in section 4.2, to calculate the GW intensity in case of GW production from domain walls. The selected values of $V_{\rm{bias}}$ for all the BPs are mentioned in Table \ref{t2}. 
 In Figure \ref{fig:2} we show the variation of the parameter $S_3/T$ with temperature $T$ for BP 3. We obtained similar variations for other BPs but for the purpose of demonstration we only present here the results for BP 3. By calculating the slope of the $(S_3/T\hspace{1mm} {\rm{vs}}\hspace{1mm} T)$ graph we compute the value of the parameter $\beta$ using Eq. \ref{eq:67}. In Figure \ref{fig:2} we also draw a horizontal line (red color) at $S_3/T=140$ to indicate the onset of bubble nucleation. 

\begin{figure}[H]
\centering
\includegraphics[width=8.55cm,height=8.55cm]{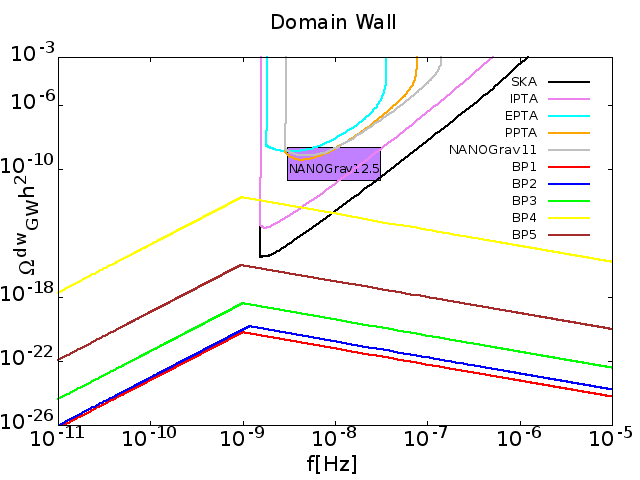}
\includegraphics[width=8.55cm,height=8.55cm]{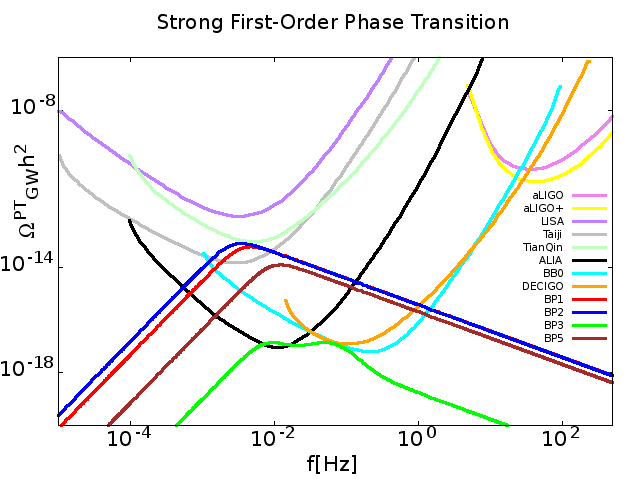}
\caption{The GW intensity versus the frequency. Left panel: comparison with the sensitivity curves of pulsar timing arrays such as SKA, IPTA, EPTA, PPTA, NANOGrav11 and NANOGrav12.5 (purple coloured rectangular box) from the contribution of the annihilation of domain walls. Right panel: comparison with the sensitivity curves of future GW detectors such as ALIA, BBO, DECIGO, LISA, TianQin, Taiji, aLIGO and aLIGO+ from the contribution of strong first-order phase transition using the same BPs.}
\label{fig:3}
\end{figure}

Finally in Figure \ref{fig:3} we plot the variation of GW intensity as a function of frequency. In the left panel of Figure \ref{fig:3} we show the GW intensity from domain walls for our selected five BPs whereas in the right panel of Figure \ref{fig:3} the GW intensity from phase transition for the same BPs are plotted. We perform a direct comparison between our calculated model-dependent GW intensity with the sensitivity plots \cite{Schmitz:2020syl} of PTAs such as SKA, IPTA, EPTA, PPTA, NANOGrav11 and NANOGrav12.5 for the domain wall case and found that the calculations using BP 4 lie within the sensitivity curves of SKA and IPTA. We note that calculations using BP 4 yield higher GW intensity compared to the results when other BPs are used when GWs from domain walls are considered but the GWs from first-order phase transition cannot be obtained if BP 4 is used for the calculation. We also find that the VEVs $v_1$, $v_2$ play important roles for GW intensities from domain walls. From Tables \ref{t1}, \ref{t4} and Figure \ref{fig:3} we may conclude that higher the values of VEVs, higher are the  GW intensities for domain wall scenario. 

We also compare our calculated GW spectrum from SFOPT (for same BPs) with the future space-based and ground-based gravitational wave detectors such as ALIA, BBO, DECIGO, LISA, TianQin, Taiji, aLIGO and aLIGO+ and find that GW intensities obtained using BP 1 lies within the sensitivity curves \footnote[1]{In this work we consider the power-law-integrated sensitivity curves \cite{Thrane:2013oya, {Dev:2019njv}}. For an alternative approach to represent the sensitivity curves see \cite{Schmitz:2020syl, {Moore:2015}, {Alanne:2019bsm}}} of ALIA, BBO, DECIGO, Taiji and marginally lies in TianQin while for the case of BP 2 the GW intensities lie within the sensitivity curves of ALIA, BBO, DECIGO, Taiji. The GW intensities marginally lie within the sensitivity curves of ALIA, BBO, DECIGO when BP 3 is used and for BP 5 they lie within the sensitivity curves of ALIA, BBO and DECIGO. 

From the results mentioned in Table \ref{t2} - \ref{t5}, we conclude that the dominant parameter influencing GW in case of GWs from domain wall is $\sigma_{\rm{wall}}$ whereas in the case of SFOPT the dominant parameter is $\beta$.    

\section{Summary and Conclusions}

In this work, we have explored two possible production mechanisms of GWs, one is  from the annihilation of domain walls and the other is from the strong first-order phase transition in the early Universe, within the framework of a particle physics model where SM is extended by two complex scalars. We investigate the detection possibilities of these GWs at the pulsar timing arrays such as SKA, IPTA, EPTA, PPTA, NANOGrav11 and NANOGrav12.5 as well as at future space-based and ground-based gravitational wave detectors such as ALIA, BBO, DECIGO, LISA, TianQin, Taiji, aLIGO and aLIGO+. A discrete $\mathcal{Z}_2 \times \mathcal{Z}^{\prime}_2$ symmetry is imposed in the model and the two added complex scalar fields $S_1$ and $S_2$  acquire a non zero vacuum expectation value when this imposed symmetry is spontaneously broken. As the discrete symmetry is broken spontaneously, domain walls are formed which is made unstable by considering this symmetry to be approximate which is explicitly broken by introducing a bias term in the theory. After the formation of these unstable domain walls, they annihilate and produce a significant amount of GWs. We also discuss the production procedure of GWs from a strong first-order phase transition. For that, we consider a finite temperature effective potential and show that the potential induces a first-order phase transition. 
We choose five BPs to calculate the GW intensity and frequency from both the production mechanisms (annihilation of domain walls and strong first-order phase transition). The BPs are chosen in such a way that they satisfy both the theoretical constraints such as vacuum stability, perturbativity and the experimental constraints such as collider bounds. We found that the VEVs $v_1$, $v_2$ (VEVs acquired by $S_1$ and $S_2$ after spontaneous breaking of $\mathcal{Z}$ and $\mathcal{Z}^{\prime}$ respectively) play a very important role to calculate the GW intensity from the collapse of domain walls and higher values of VEVs give higher GW intensities. The peaks for the GW intensities for their production from two mechanisms, namely domain wall annihilation and strong first-order phase transition, appear in two different frequency regions. For the case of GWs from domain wall annihilation, the intensity peaks in a lower frequency regime around $\sim 10^{-9}$ Hz whereas for GWs from strong first-order phase transition, these peaks are obtained at comparatively higher frequency regime of $\sim 10^{-3} - 10^{-2}$ Hz. We also demonstrate that our calculated GWs from the annihilations of domain walls can be probed by low-frequency PTAs such as SKA and IPTA whereas the GWs which are produced from strong first-order phase transition can be accessed by comparably higher frequency detectors such as ALIA, BBO, DECIGO, TianQin and Taiji. In this work, within the framework of our proposed model we show that the simple extension of SM can explain the formation of unstable domain walls as well as the strong first-order electroweak phase transition. One can also extend this model for a particle dark matter theory by considering one of the components of the complex scalars to be a dark matter candidate and work out its phenomenology as well as the viability of such a dark matter candidate. This is for posterity.

\vspace{5mm}
\noindent {\bf Acknowledgments}

The authors would like to thank A. Dutta Banik for useful discussions. AP would like to thank Biswajit Banerjee for helping in modifying the Cosmotransition package. One of the authors (U.M.) receives her fellowship (as a graduate student leading to Ph.D. degree) grant from Council of Scientific \& Industrial Research (CSIR), Government of India as Senior Research Fellow (SRF) having the fellowship Grant No. 09/489(0106)/2017-EMR-I.

{}
\end{document}